\documentclass[preprint,12pt]{elsarticle}

\usepackage{lineno,hyperref}
\modulolinenumbers[5]

\newcounter{bla}

\journal{Optics Communications}

\usepackage[bbgreekl]{mathbbol}
\usepackage{amsfonts}
\usepackage{amsmath}
\usepackage{amssymb}
\usepackage{booktabs}
\usepackage{tabularx}
\usepackage{mathtools}
\usepackage{pgfplots}
\usepackage{listings}
\usepackage{geometry}
\usepackage{pdflscape}
\usepackage{enumerate}
\usepackage{esvect}
\usepackage[figuresright]{rotating}
\usepackage{footnote}
\usepackage{listings}
\usepackage{xcolor}
\usepackage{enumitem}
\usepackage{tocloft}
\usepackage{textcomp}
\usepackage{braket}
\usepackage{nicefrac}
\usepackage{nccmath}
\usepackage{bm}
\usepackage{import}
\usepackage{dpfloat}
\usepackage{nicefrac}
\usepackage{subcaption}
\usepackage{float}
\usepackage{afterpage}
\usepackage{multirow}
\usepackage{siunitx}
\usepackage{graphicx}
\usepackage{tikz}
\definecolor{Darkgreen}{rgb}{0,0.4,0}
\usepackage[ruled]{algorithm2e}
\graphicspath{{Figs/}}
\DeclarePairedDelimiter\abs{\lvert}{\rvert}%
\definecolor{light-gray}{gray}{0.95}

\usepackage{graphicx}
\newcommand{\CC}{%
    {\settoheight{\dimen0}{C}C\kern-.05em \resizebox{!}{\dimen0}{\raisebox{\depth}{++}}}}
\newcommand{\CCC}{%
    {\settoheight{\dimen0}{C}C/C\kern-.05em \resizebox{!}{\dimen0}{\raisebox{\depth}{++}}}}
\newcommand{\CS}{%
    {\settoheight{\dimen0}{C}C\kern-.05em \resizebox{!}{\dimen0}{\raisebox{\depth}{\#}}}}

\definecolor{mygray}{rgb}{0.4,0.4,0.4}
\definecolor{mygreen}{rgb}{0,0.8,0.6}
\definecolor{myorange}{rgb}{1.0,0.4,0}

\begin{document}
    \suppressfloats 
    
     \begin{frontmatter}
        
        \title{Improving performance of single-pass real-time holographic projection}
        
        \author[mymainaddress]{Peter J. Christopher\corref{mycorrespondingauthor}}
        \cortext[mycorrespondingauthor]{Corresponding author}
        \ead{pjc209@cam.ac.uk}
        \ead[url]{www.peterjchristopher.me.uk}
        
        \author{Ralf Mouthaan}
        
        \author{Vamsee Bheemireddy}
        
        \author{Timothy D. Wilkinson}
        
        \address[mymainaddress]{Centre of Molecular Materials, Photonics and Electronics, University of Cambridge}
        
        \begin{abstract}
            This work describes a novel approach to time-multiplexed holographic projection on binary phase devices. Unlike other time-multiplexed algorithms where each frame is the inverse transform of independently modified target images, Single-Transform Time-Multiplexed (STTM) hologram generation produces multiple sub-frames from a single inverse transform. Uniformly spacing complex rotations on the diffraction field then allows the emulation of devices containing $2^N$ modulation levels on binary devices by using $N$ sub-frames. In comparison to One-Step Phase Retrieval (OSPR), STTM produces lower mean squared error for up to $N=5$ than the equivalent number of OSPR sub-frames with a generation time of $\nicefrac{1}{N}$ of the equivalent OSPR frame.  A mathematical justification of the STTM approach is presented and a hybrid approach is introduced allowing STTM to be used in conjunction with OSPR in order to combine performance benefits.
        \end{abstract}
        
        
    \end{frontmatter}

	\section{Introduction}
	
    Holograms are widely used in a variety of fields including imaging~\cite{Svoboda2013, Frauel2006, Sheen2001}, displays~\cite{Maimone2017,Yamada2018}, lithography~\cite{Turberfield2000, Purvis2014} and optical manipulation~\cite{Grier06,Melville2003,Grieve2009}. Computer-generated holograms are made possible by devices known as spatial light modulators (SLMs) that allow either the amplitude or phase of light incident on the device to be independently controlled and a diffraction pattern to be created. The task of projecting a desired field pattern then becomes that of finding an appropriate modulation that takes into account the practical constraints imposed by the SLM. 
		
	Established iterative search algorithms such as direct search and simulated annealing can yield very low mean-squared errors, making them well-suited to applications where the complexity of the desired diffraction field is low or where a handful of high-quality algorithms can be calculated offline. These have been successfully used in holographic optical switches \cite{battig92, Georgiou2008}, for optical fibre mode excitation \cite{carpenter2010graphics} and in optical security systems \cite{Abookasis06}, but are not appropriate for online, real-time calculation as the computational demands can be very high even for GPUs and FPGAs. Similarly, phase retrieval algorithms such as the Gerchberg-Saxton algorithm have been shown to work well for multi-level phase SLMs, but tend to diverge for binary SLMs. Importantly, none of these algorithms are particularly well-suited for generating video holograms for visual consumption, in that they do not accommodate for the unique response of the human eye, are not necessarily guaranteed to converge in all cases, and  are not computationally efficient enough to generate high-resolution holograms on the fly and in real time.
		
	In 2006, Cable \& Buckley introduced One-Step Phase Retrieval (OSPR) ~\cite{buckley200870, buckley2011real, buckley2011computer, cable200453, Buckley06thesis, Buckley06}, a computationally efficient hologram generation algorithm specifically designed with video display applications in mind and capable of generating subjectively pleasing holograms in real-time. The algorithm produces time-multiplexed sub-frames but is otherwise distinct from other holographic time-multiplexing techniques that, for example, aim to reduce speckle \cite{Amako1995,Takaki2011,Liu2017} or produce full-colour holograms \cite{Shimobaba2007, Shimobaba2003, Oikawa2011}. In this paper we briefly re-introduce OSPR  before presenting our novel approach to real-time holographic video projection, the Single-Transform Time-Multiplexed (STTM) algorithm. 
    
    \begin{figure}[tb]
        \centering
        {\includegraphics[trim={0 0 0 0},width=0.25\linewidth,page=1]{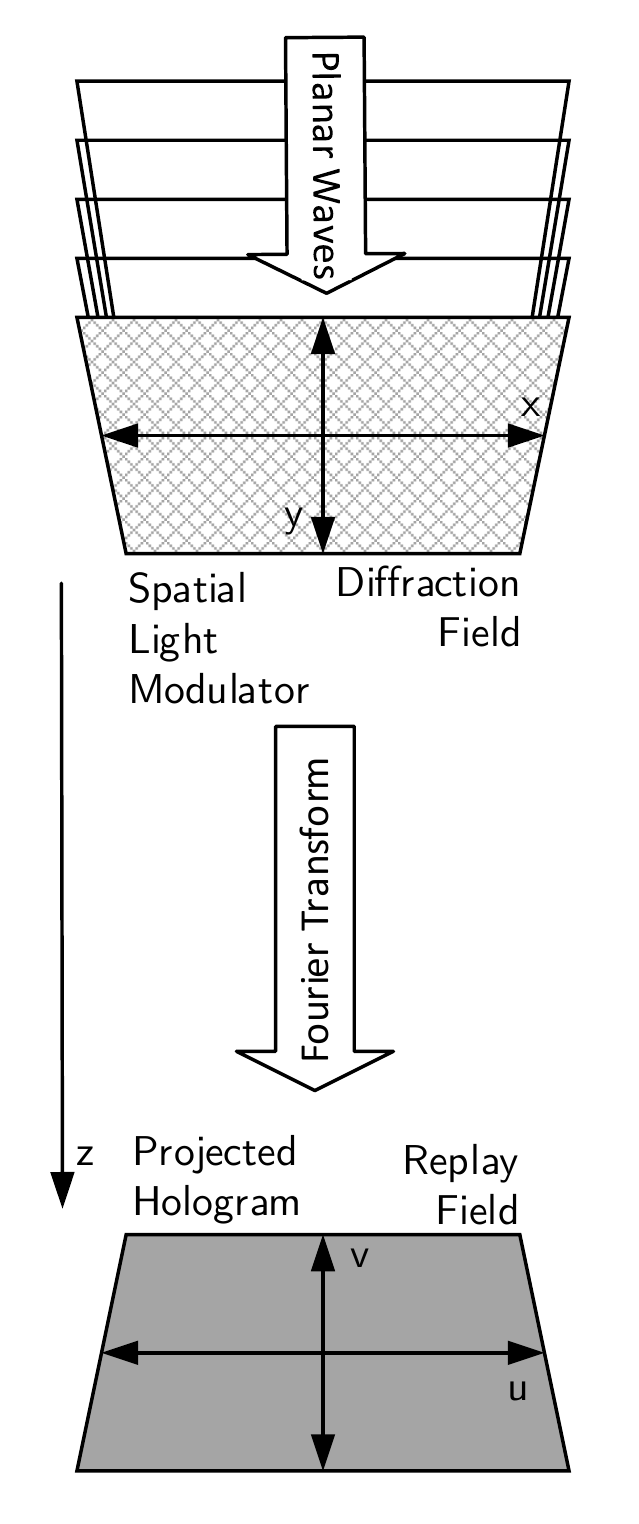}}
        \hspace{0.01\linewidth}
        {\includegraphics[trim={0 0 0 0},width=0.295\linewidth,page=1]{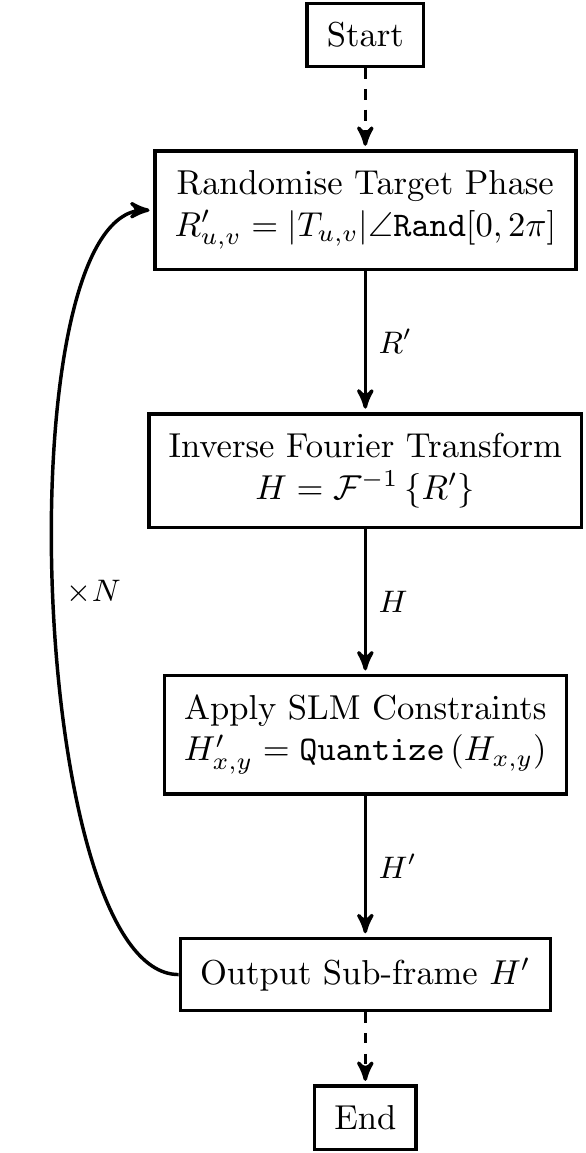}}
        \hspace{0.01\linewidth}
        {\includegraphics[trim={0 0 0 0},width=0.35\linewidth,page=1]{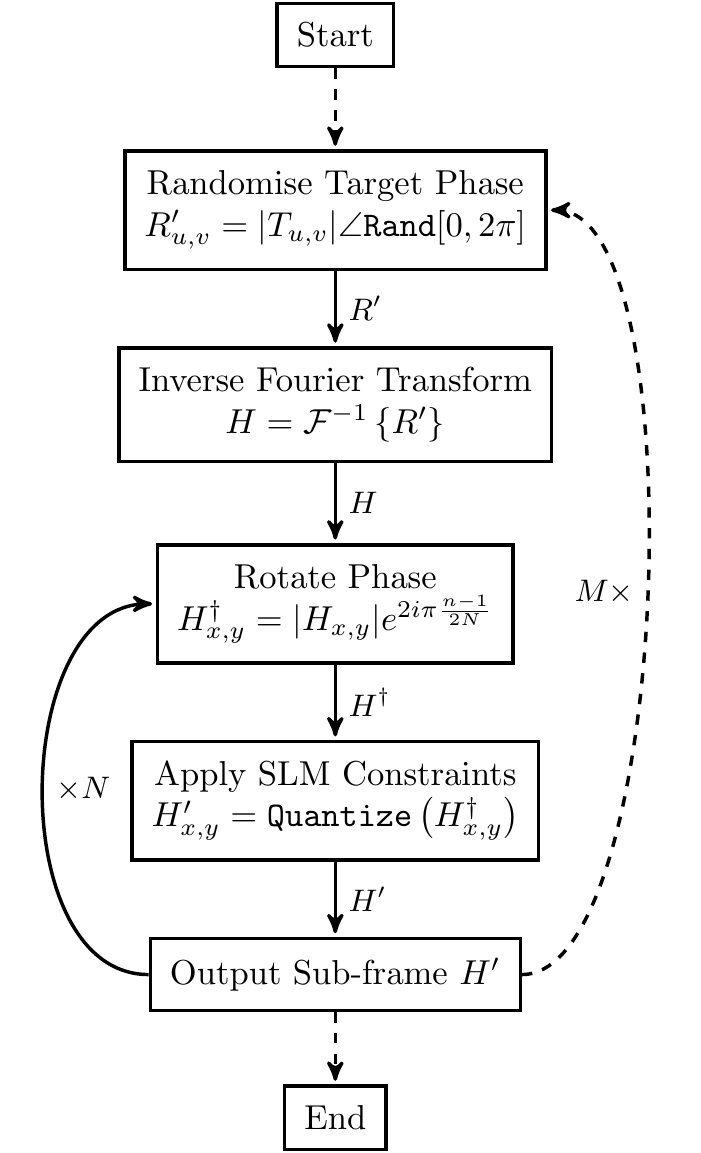}}
        \caption{Coordinate systems used in describing a hologram (left), One-Step Phase-Retrieval algorithm (centre) and Single-Transform Time-Multiplexed algorithm (right)}
        \label{fig:algs}
    \end{figure}
		
\section{Algorithms}
    
\subsection{One-Step Phase-Retrieval}
		
In 2006, Cable \& Buckley made several observations regarding the human eye's response to light:
		
		\begin{itemize}
			\item The eye finds images with low noise variance, as opposed to low bias, pleasing;
			\item The eye responds to the intensity of incident light, but is agnostic to the phase of incident light;
			\item The eye has a response time of approximately 40~ms. Within this window, the impulse response of the eye can be modeled as a square pulse where multiple received stimuli can simply be averaged with equal weighting;
		\end{itemize}
		
		Consequently, Cable \& Buckley proposed efficiently generating many low-quality holograms to be presented to the viewer in rapid succession. For example, 24 sub-frames making up a single frame are to be presented to the viewer within 1/60th of a second using a ferroelectric liquid crystal SLM (capable of operating at 1440~Hz). The slower response time of the human eye would smooth out these low-quality sub-frames, giving the overall impression of a much higher quality image.
		
		The efficient algorithm proposed, coined One-Step Phase Retrieval (OSPR) is shown in Fig. \ref{fig:algs}~(middle). Briefly, light diffracted by an aperture $f(x,y)$ and projected onto a flat two-dimensional surface creates a field pattern $F(u,v)$ as shown in Figure~\ref{fig:algs}~(left). In so-called Fraunhofer systems, where the projection is into the far-field, the projected pattern is the Fourier transform of the aperture $F(u,v) = \mathcal{F}\{f(x,y)\}$, where $x$, $y$ are the spatial coordinates of the diffraction field $u$ and $v$ are the spatial coordinates of the replay field. For regularly sampled $f(x,y)$ and $F(u,v)$ this transform can be efficiently calculated using the discrete Fourier transform of Eq. \ref{DFT}. The phase of the target field pattern is first randomised before the DFT is taken to yield an appropriate aperture function. The constraints of the SLM are then applied, and in doing so a low-quality hologram that will act as a single sub-frame is obtained. The process is repeated $N$ times where $N$ is the desired number of sub-frames and $N_x$ and $N_y$ are the number of pixels on the $x$ and $y$ axes respectively.
		
		\begin{equation} \label{DFT}
		F(u,v) = \frac{1}{\sqrt{N_xN_y}} \sum_{N_x} \sum_{N_y} f(x,y) e^{\frac{ux}{N_x} + \frac{vy}{N_y}}
		\end{equation}
        
		Of particular note is the initial phase randomisation performed at the beginning of each iteration. This ensures an independent hologram is obtained after each iteration, but also serves to smooth the Power Spectral Density~(PSD) of the target image and to reduce edge enhancement. The randomisation of the target image phase in this manner is only applicable in phase insensitive applications where only the replay field intensities are of interest. 
    
\subsection{Single-Transform Time-Multiplexed}
    
    In this work we introduce an alternative time-multiplexed algorithm to OSPR known as Single-Transform Time-Multiplexed (STTM) hologram generation.
    
    OSPR performs $N$ independent DFT operations. This is  computationally expensive in real-time applications where 1000s of sub-frames may require processing every second. Instead, an alternative algorithm is proposed. The phase of the target field pattern is first randomised and the inverse DFT is taken to obtain the diffraction aperture, in a similar manner to many other algorithms. At this stage the complex values of the obtained hologram are rotated through a phase angle of $2\pi \frac{n-1}{N}$ before the SLM constraints are applied to obtain the sub-frame. This is repeated for $n = 1..N$. As the DFT magnitude is invariant under rotation in the complex plane, this does not effect the far-field amplitudes. This change in approach significantly reduces the computational load when compared to OSPR. It is noted that the STTM algorithm easily lends itself to parallel execution.
		
\subsection{Hybrid STTM}

Finally, a hybrid variant is also proposed. This proceeds in a manner similar to the STTM algorithm, except that it is periodically restarted, as illustrated in Fig. \ref{fig:algs}~(right). Consequently, the replay field is phase-randomised and the inverse DFT taken $M$ times, and the phase angle of the hologram is shifted by  $2\pi \frac{n-1}{N}$ for $n=1..N$ times, to yield $M$ sets of $N$ sub-frames.
    
\section{Results}
    
\begin{figure}[tbhp]
        \centering
        {\includegraphics[trim={0 0 0 0},width=\linewidth,page=1]{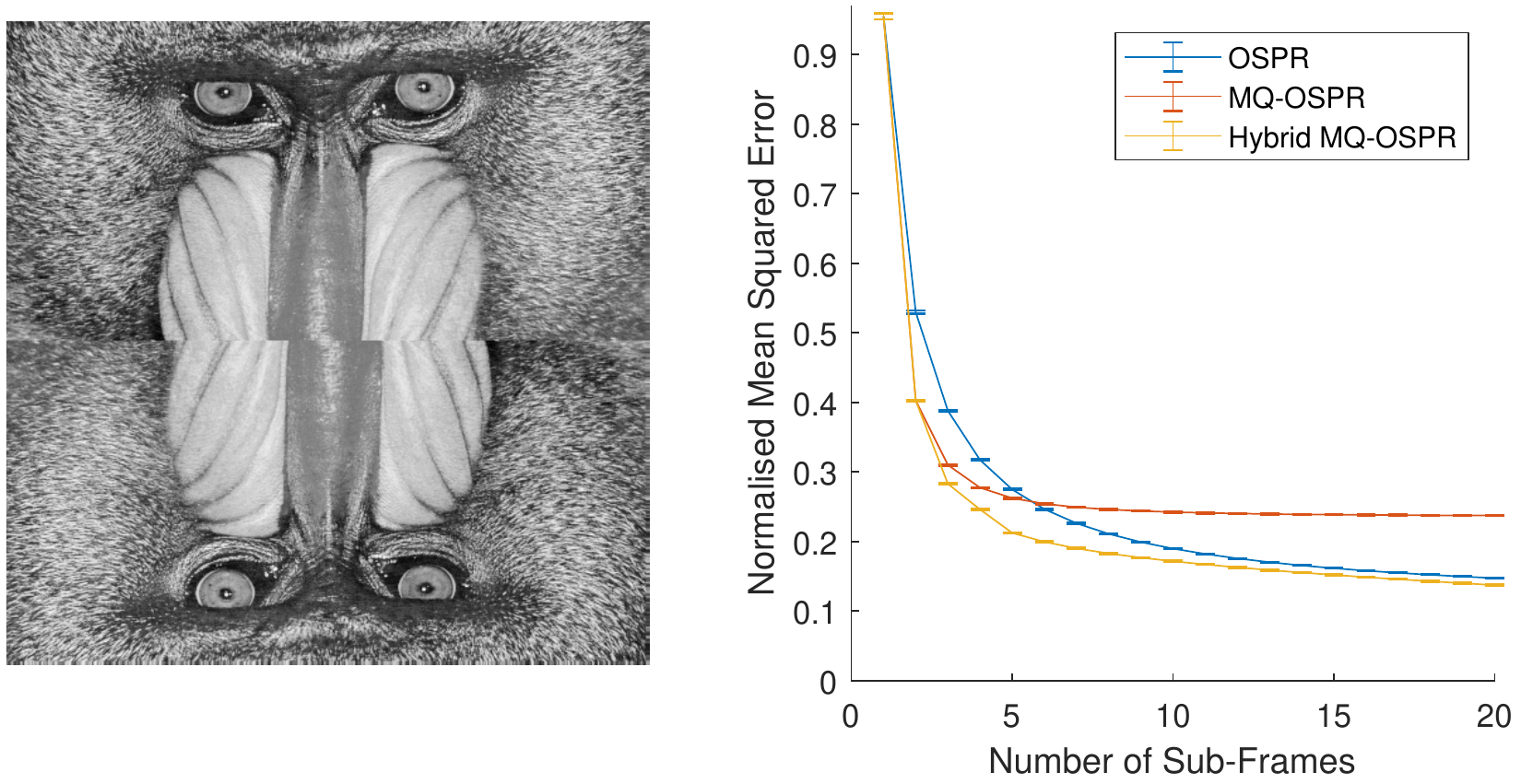}}
        \caption{Left: 512x512 Mandrill test image with induced rotational symmetry. Right: Time-averaged errors for different OSPR variants run on the Mandrill image (left). Values are taken as being the mean of $100$ independent runs with error bars showing one standard deviation.}
        \label{fig:mqospr_001}
    \end{figure}

    Figure~\ref{fig:mqospr_001} shows the phase-insensitive mean-squared error (MSE) convergence of OSPR as well as STTM and hybrid STTM, as calculated from Eq. \ref{msepi}. This error metric is adopted as it encompasses both bias and variance errors. Values are taken as being the mean of $100$ independent runs with error bars showing one standard deviation. The $512\times512$ pixel \textit{Mandrill} test image with artificially induced rotational symmetry of Fig. \ref{fig:mqospr_001} (left) is used for the target. 
    
\begin{figure}[tbhp]
    	\centering
    	{\includegraphics[trim={0 0 0 0},width=1.0\linewidth,page=1]{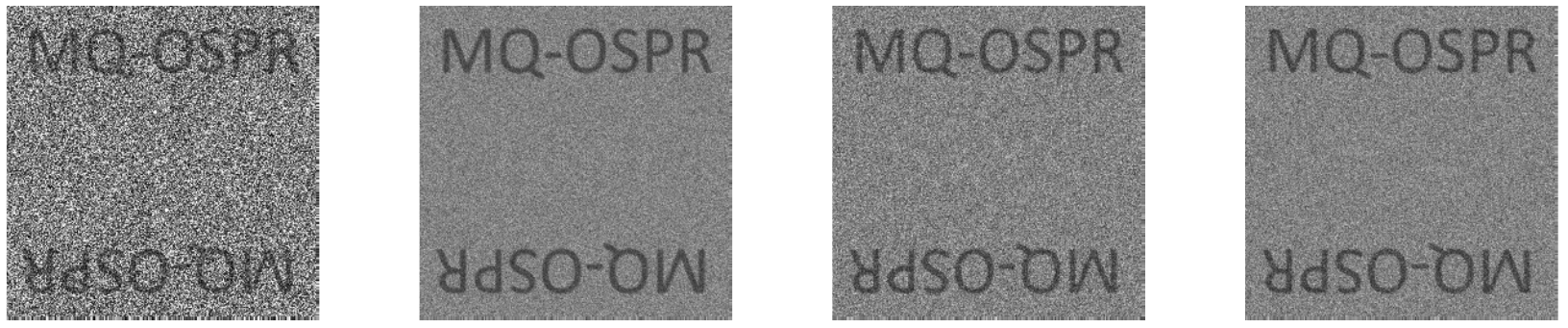}}
    	\caption{Comparison of OSPR and STTM showing a single frame of OSPR (far left), 12 frames of OSPR (centre left), 12 frames of STTM (centre right) and 3 sets of 4 STTM sub-frames (far right). Target image is $256 \times 256$ pixels and the simulated device binary phase.}
    	\label{fig:mqospr_004}
    \end{figure}

    A comparison of the computer-generated replay fields generated by OSPR, STTM and hybrid STTM is shown in Figure~\ref{fig:mqospr_004}. The shown images correspond to the equally-weighted sum of the obtained sub-frames to mimic the impulse response of the human eye. Low resolutions are provided to ease comparison. The STTM image (centre right) had a mean-squared error less than 20\% greater than the OSPR generated frames (centre right) and was generated in less than 10\% of the time. The combined hybrid frame (far right) was generated from 3 sets of 4 STTM sub-frames and had an error 5\% less than the OSPR equivalent and was generated in 30\% of the time. The relative speed up becomes even more significant at higher resolutions where the FFT step takes up a greater percentage of the performance impact.
    
\begin{figure}[tbhp]
        \centering
        {\includegraphics[width=1.0\textwidth]{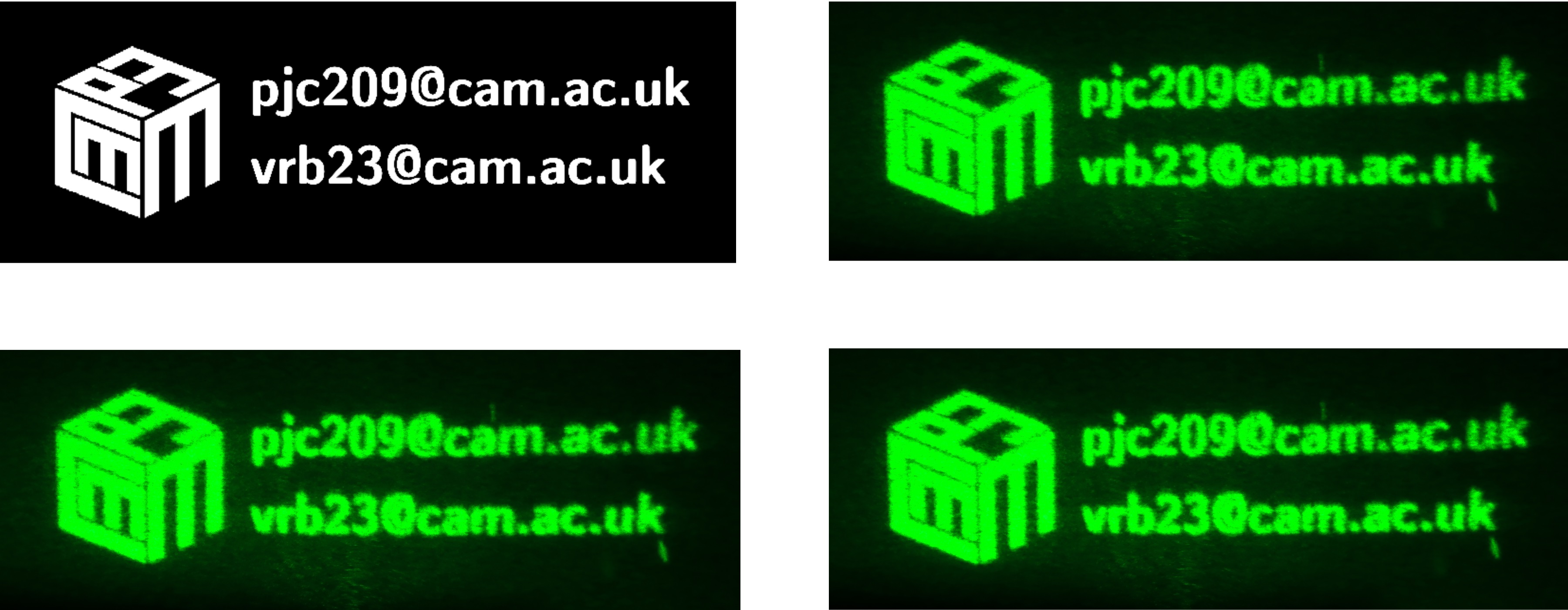}}
        \caption{Target image (top left) with 24 sub-frames shown on a $1024 \times 1024$ pixel binary phase projector with a ball lens. The top half of the holographic replay fields are shown for 24 frames of OSPR (bottom left), 24 frames of STTM (top right) and 4 sets of 6 frames of STTM (bottom right). The STTM and hybrid frames are generated in $< 5\%$ and $< 20\%$ respectively of the time with little visual difference. Captured using a Canon 5D Mark III with a 24-105mm lens and a $\nicefrac{1}{60}$ second exposure.}
        \label{fig:test}
    \end{figure}

	In order to experimentally confirm these computational results, we generated $3$ $1024 \times 1024$ binary phase holograms using the target image shown top left in Figure~\ref{fig:test}. The first of these was generated using 24 sub-frames of OSPR, Figure~\ref{fig:test} (bottom left), the second using 24 sub-frames of STTM, Figure~\ref{fig:test} (top right), and the third using 4 sets of 6 sub-frames of STTM, Figure~\ref{fig:test} (bottom right). The STTM and hybrid frames are generated in $< 5\%$ and $< 20\%$ respectively of the time taken to generate the OSPR frame. The SLM used was a ferroelectric $1024\times 1024$ device from Forth Dimension Displays. The configuration used was discussed previously in~\cite{freeman2010visor2}.
    
\section{Discussion}

	Inspection of Fig. \ref{fig:test} shows that, once projected using a high-speed ferroelectric SLM, there is little to no visual difference between the replay fields, whereas inspection of Fig. \ref{fig:mqospr_004} reveals that the STTM algorithm produces holograms that are only marginally inferior to the original OSPR algorithm, but at a fraction of the computational cost. In practical applications it would be possible to present many more STTM sub-frames within a given time period but the MSE improvements observed would be ever diminishing. 

	Inspection of Fig. \ref{fig:mqospr_001} reveals that STTM actually out-performs OSPR for the first few sub-frames. It is this observation that motivated the development of the hybrid-OSPR algorithm, which is in turn seen to converge to the lowest MSE.
    
    In order to provide a relationship for the expected MSE reduction we make a three-stage argument. Firstly, we show that the expected distribution of diffraction field magnitudes for any distribution of replay field magnitudes with uniformly distributed phase must follow a Rayleigh distribution. Secondly, we show that for a Mean Squared Error (MSE) estimator the expected error of modifying a single pixel is proportional to the square of the distance moved. Thirdly, we show that using $N$ binary quantised subframes are in fact equivalent to a single frame displayed on a device with $2N$ modulation levels. Finally we combine these relationship to provide an analytical relationship for error reduction against number of iterations for STTM. This shows that the convergent error is expected to be $\approx 26\%$ of the first iteration error.
    
    \subsection{Expected distribution of diffraction field values} \label{subsec1}
    
    The first step we must follow is to develop a theory for the expected distribution of diffraction field values. If we consider a replay field with a distribution of amplitudes $\mathbb{R}_r$ and a distribution of phases $\mathbb{\Phi}_r$. $\mathbb{R}_r$ is assumed to be an arbitrary distribution with variance $\sigma_r^2$ which we choose to normalise to $1$. $\mathbb{\Phi}_r$ is assumed to be uniformly distributed in the interval $\left[ 0,2\pi \right)$ and independent of $\mathbb{R}_r$.
    
    \begin{equation} \label{Zr}
    \mathbb{C}_r=\mathbb{R}_r e^{i\mathbb{\Phi}_r}
    \end{equation}
    
		The diffraction field is related to the replay field by the inverse DFT. The distribution of values $\mathbb{C}_d$ taken on by a given diffraction field pixel is hence given by
    
    \begin{equation}
    \mathbb{C}_d=\frac{1}{\sqrt{N_uN_v}} \sum_{N_u} \sum_{N_v} \mathbb{R}_r e^{ i \mathbb{\Phi}_r} e^{2\pi i \left(\frac{u x}{N_u} + \frac{v y}{N_v}\right)}
    \end{equation}
    
    where $N_u$ and $N_v$ here represent the number of pixels on the $u$ and $v$ axes respectively. This summation is over a set of vector variables and as such tends towards a Rayleigh distribution of the form of Eq. \ref{ampdist} for large $N_uN_v$.
    
    \begin{equation} \label{ampdist}
    p(r)=2r e^{ -r^2 }
    \end{equation}
    
\begin{figure}[htbp]
    	\centering
    	{\includegraphics[trim={0 0 0 0},width=\linewidth,page=1]{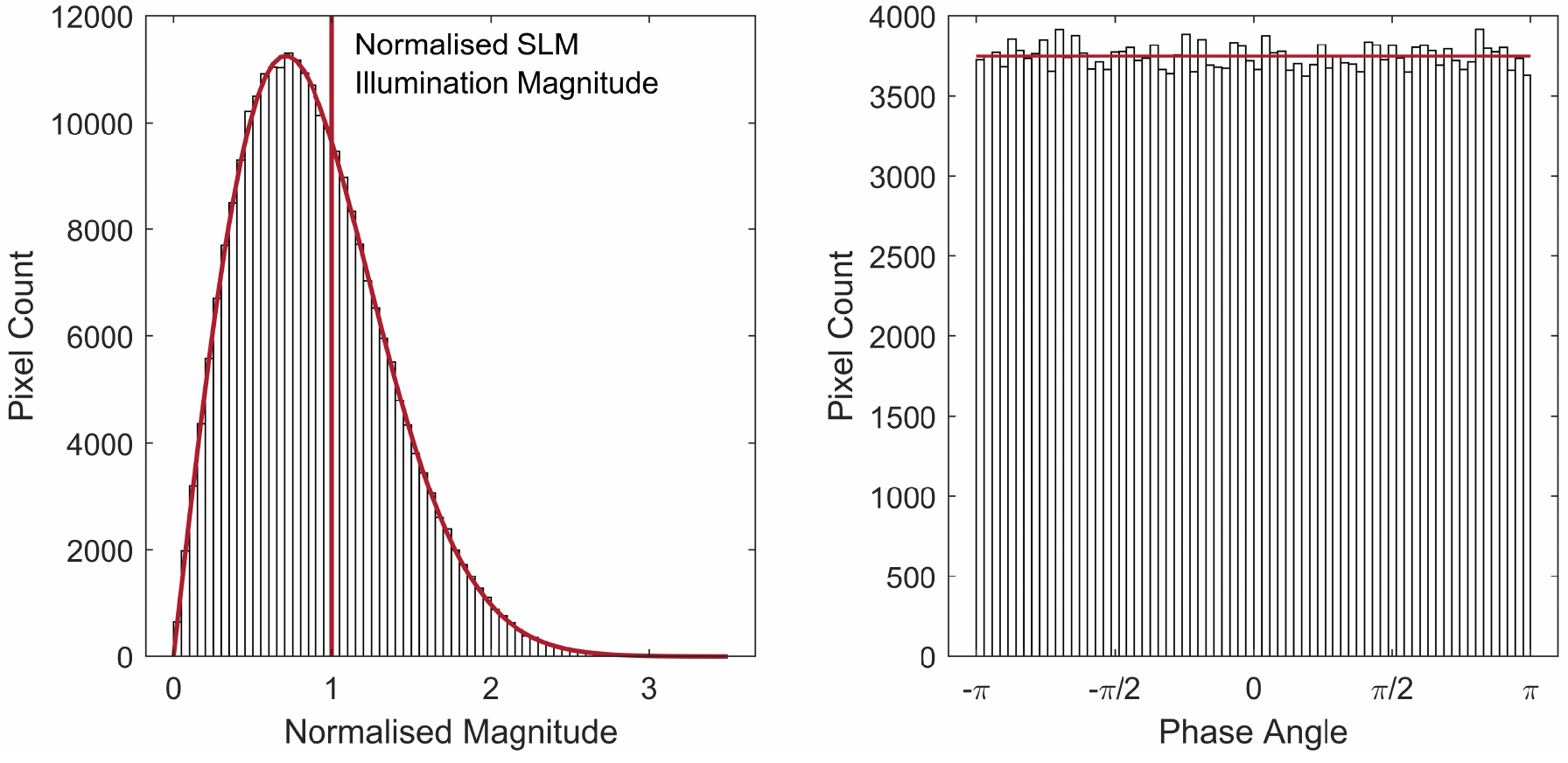}}
    	\caption{Spread of magnitudes (left) and angles (right) of the inverse Fourier transform of the phase randomised $512\times512$ \textit{Mandrill} test image. The expected distribution is shown in red.}
    	\label{fig:hist1}
    \end{figure}

    This result relies on the central limit theorem, and is consequently only valid for large $N_uN_v$. As the only further restriction we place on the replay field pixel magnitudes was that they be normalised to unit variance, this formula applies to any expected magnitude distribution in the replay field, not just a uniform distribution. For example, the expected diffraction field magnitudes and phases of the phase randomised $512\times512$ \textit{Mandrill} test image is shown in Figure~\ref{fig:hist1}. 
    
	\subsection{Expected error as a function of quantisation change} \label{subsec2}
	
	The per-pixel phase insensitive MSE formula is given as a function of the target image $T$ and generated replay field $R$
	
	\begin{equation} \label{msepi}
	E_{\text{MSE}}(T,R) = \frac{1}{N_x N_y}\sum_{u=0}^{v=N_x-1}\sum_{y=0}^{y=N_y-1} \left[\abs{T_{u,v}} -  \abs{R_{u,v}}\right]^2.
	\end{equation}
	
	\begin{figure}[htbp]
		\centering
		{\includegraphics[trim={0 0 0 0},width=\linewidth,page=1]{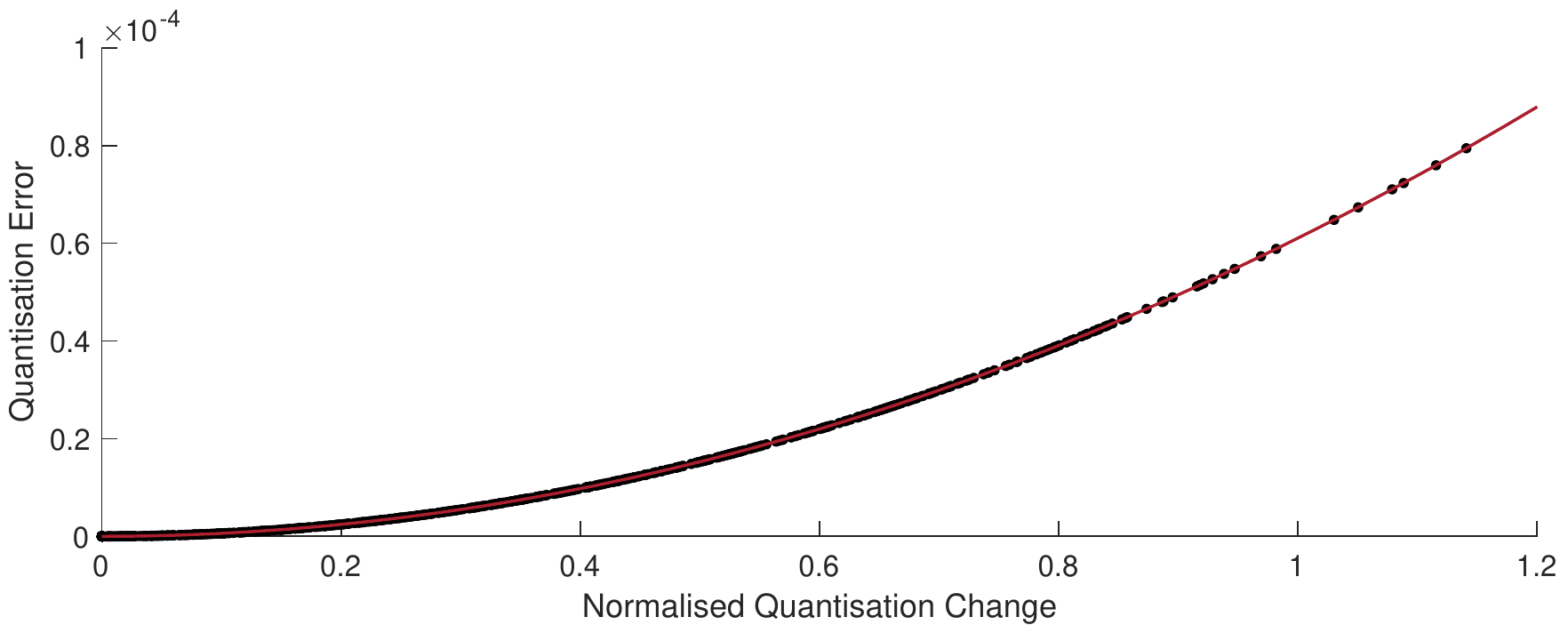}}
		\caption{Scatter plot of pixel value changes during quantisation of a continuous phase hologram against resultant error as well as the expected trend line. Test image used is a 512 $\times$ 512 pixel version of \textit{Mandrill}. }
		\label{fig:hist3}
	\end{figure}
	
	The change in the a replay field pixel $\Delta R_{u,v}$ due to a change in a diffraction field pixel $\Delta H{x,y}$ is derived from the DFT identity off Eq. \ref{DFT}
	
	\begin{equation}
	\Delta R_{u,v} = \frac{1}{\sqrt{N_xN_y}} \Delta H_{x,y} e^{\frac{ux}{N_x} + \frac{vy}{N_y}}
	\end{equation}.
	
	Inserting this into Eq. \ref{msepi} and performing the summation gives an expression for the change in MSE $\Delta E_{MSE}$ due to an altered hologram pixel where $C_{\text{MSE}}$ is a constant in the range $\left[0,1\right)$. This relationship is also shown in Fig. \ref{fig:hist3}.
	
	\begin{equation} \label{mseps1}
	\Delta E_{MSE}=C_{MSE}\frac{\abs{\Delta H_{x,y}}^2}{N_x N_y}
	\end{equation}	
    
    The constant $C_{MSE}$ will decrease as further pixels are quantised and correlations are introduced between pixels. The analysis below depends only on the ratio between errors and the constant $C_{MSE}$ will cancel.
	
	\subsection{Multi-frame equivalence to multi-level quantisation} \label{subsec3}
	
\begin{figure}[htbp]
        \centering
        {\includegraphics[trim={0 0 0 0},width=0.41\linewidth,page=1]{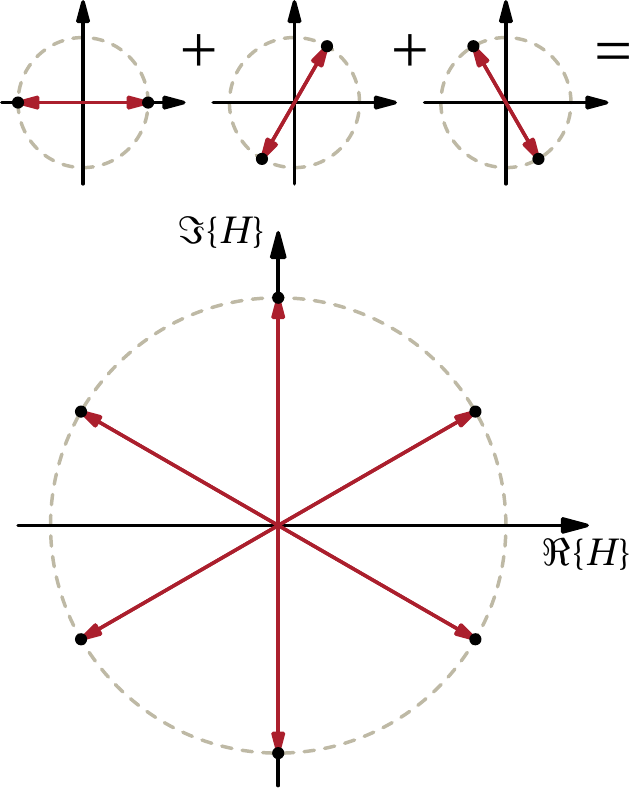}}
        \hspace{0.02\linewidth}
        {\includegraphics[trim={0 0 0 0},width=0.55\linewidth,page=1]{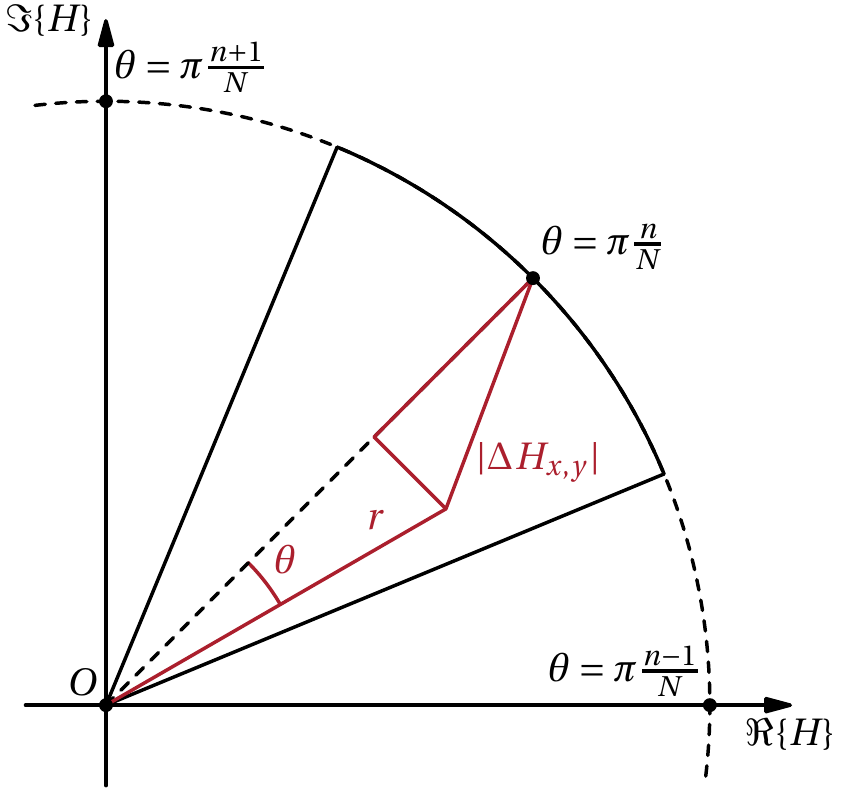}}
        \caption{Left: The sum of the binary phase quantisations of a hologram rotated through $0^{\circ}$, $60^{\circ}$ and $120^{\circ}$ is equivalent to a single hologram quantised on a six-level device. Right: Geometry of quantising and modulating a pixel.}
        \label{fig:hist4}
    \end{figure}

	By treating the time-multiplexed hologram-subframes as a linear addition of intensities it can be seen that summing the binary phase quantisation of a hologram and the binary phase quantisation of the same hologram rotated by $60^{\circ}$ and $120^{\circ}$ is the equivalent of the six phase quantisation of the hologram, as illustrated on the left side of Fig. \ref{fig:hist4}. More generally, the projection of $N$ STTM frames quantised on an SLM with $M$ levels is equivalent to projecting a single frame quantised on an SLM with $NM$ levels. This leads in the limit as $N \rightarrow \infty$ to a continuously modulated phase device.
	
	\begin{equation}
	\sum_{n=1}^{N}\underset{\scriptscriptstyle \text{binary phase}}{\texttt{Quantise}}\left[He^{2 i \pi \frac{n-1}{2N}}\right] = N\times\underset{\scriptscriptstyle 2N \text{phase levels}}{\texttt{Quantise}}\left[He^{2 i \pi \frac{1}{4N}}\right]
	\end{equation}
    
	Considering the display of $N$ binary hologram subframes to be equivalent to a single subframe on an $N$-level devices allows any a given pixel value of the hologram $H_{x,y}$ can be written in terms of amplitude $r$ and angle $\theta$ from the nearest virtual modulation level where $\theta$ where $\theta$ is in the range $\left[-\frac{\pi}{2N},\frac{\pi}{2N}\right)$ as shown in Figure~\ref{fig:hist4} (right). The distance $\abs{\Delta H_{x,y}}$ between $H_{x,y}$ and the nearest device level is then given by 
    
    \begin{equation} \label{statstart}
    \abs{\Delta H_{x,y}}=\sqrt{\left(1-r\cos{\theta}\right)^2+\left(r\sin{\theta}\right)^2}=\sqrt{1-2r\cos{\theta}+r^2}
    \end{equation}
		
\subsection{Combination} \label{combination}

	Section~\ref{subsec1} gave the probability distributions of the magnitude and phase of the diffraction field. Assuming the distributions to be independent of each other, these can be combined with the error introduced by quantising a single pixel, calculated in Section~\ref{subsec2}, to give the expected value of the error due to quantising the entire hologram on an $N$-level phase modulator. It is noted that the integral is performed over the region of the argand diagram for which pixel values map onto a virtual modulation level - this treatment is justified given the conclusions drawn in Section~\ref{subsec3}.
	
	\begin{equation}
	E_{\text{MSE,tot}} = \int_{r=0}^{\infty}\int_{\theta=0}^{2\pi/N}p(r)p(\theta) \Delta E_{MSE}(r,\theta)drd\theta \\
	\end{equation}
	
	Eqs.~\ref{ampdist}, \ref{mseps1} and \ref{statstart} are then substituted in, and $p(\theta)$ is assumed to be a uniform distribution that integrates to unity. This then gives the following expression for the total error.
    
	\begin{align}
	E_{\text{MSE,tot}}
	&=\frac{C_{\text{MSE}}}{N_x N_y}\int_{0}^{\infty}2 r e^{ -r^2}\int_{-\frac{\pi}{2N}}^{\frac{\pi}{2N}} \frac{1}{2\pi}(1-2r\cos{\theta}+r^2)  d\theta dr \nonumber \\
    &=\frac{C_{\text{MSE}}}{N_x N_y}\left(2\pi-2N\sqrt{\pi}\sin{\left(\frac{\pi}{2N}\right)}\right)
	\end{align}
	
	Normalising to the single-frame case and considering the limit as $N \to \infty$ 
    
    \begin{equation}
    E_{\text{MSE,tot},\infty}
    =E_{\text{MSE,tot,1}}\frac{\pi-\frac{1}{2}\pi^{\frac{3}{2}}}{\pi-\sqrt{\pi}}\approx0.2611E_{\text{MSE,tot,1}}
    \end{equation}
    
    which is in agreement with the observed behaviour.
    
	\section{Conclusion}
	
    The STTM hologram generation algorithm has been introduced, which allows subjectively pleasing time-multiplexed holograms to be generated in real-time. Generation times have been shown to be an order of magnitude faster than competing time-multiplex algorithms such as OSPR, with an associated degradation in performance of less than 20\%. A mathematical rationale has been given for the performance of the STTM algorithm and a hybrid STTM/OSPR algorithm has been developed that combines the advantages of each of the two approaches. The speed-up offered by the STTM algorithm offers the potential for higher-resolution, higher-framerate and more cost effective holographic displays.
	
    \section*{Acknowledgements}
    
    The authors would like to thank the Engineering and Physical Sciences Research Council (EP/L016567/1 and EP/L015455/1) for financial support during the period of this research.
    
	\bibliography{references}
	
\end{document}